\documentclass[reprint, twocolumn, superscriptaddress,prb]{revtex4-2}
\usepackage{bm, amsmath, amsfonts, amssymb, braket}
\usepackage{subfigure}
\usepackage{color}
\usepackage{graphicx}
\usepackage{slashed}
\usepackage{etoolbox}
\apptocmd{\thebibliography}{\raggedright}{}{}

\usepackage{dcolumn} % Align table columns on decimal point

\usepackage{rotating} % for 90 degree rotated table and fig

\usepackage{ulem}
\usepackage{xcolor}
\DeclareRobustCommand{\erase}{\bgroup\markoverwith{\textcolor{red}{\rule[.5ex]{2pt}{0.4pt}}}\ULon}

\begin{document}
\title{Steady-state skin effect in bosonic topological edge states under parametric driving}

\author{Nobuyuki Okuma}
\email{okuma@hosi.phys.s.u-tokyo.ac.jp}
\affiliation{
 Graduate School of Engineering, Kyushu Institute of Technology, Kitakyushu 804-8550, Japan
}

\date{\today}
\begin{abstract}
Non-Hermitian systems have attracted significant theoretical interest due to their extreme properties. However, realizations have mostly been limited to classical applications or artificial setups. 
In this study, we focus on the quantum nature inherent in bosonic Bogoliubov–de Gennes (BdG) systems, which from the perspective of spectral theory corresponds to non-Hermiticity. Based on this insight, we propose a steady-state skin effect in quantum condensed matter utilizing such BdG non-Hermiticity. Specifically, we introduce BdG quantum terms arising from parametric pumping to the edge states of an underlying bosonic Hermitian Chern insulator, thereby realizing non-Hermiticity without dissipation. This system design has the advantage of being largely independent of microscopic model details. Through analysis using non-equilibrium Green’s functions, we find that under open boundary conditions, a steady state exhibiting the non-Hermitian skin effect is realized. The pronounced corner particle accumulation observed in this steady state shows quadrature anisotropy, which manifests the bosonic quantum nature.
Our results bridge the gap between the fascinating mathematics of non-Hermitian matrices and practical quantum physical systems.

\end{abstract}

\maketitle
\section{Introduction}
\begin{figure}[]
\begin{center}
 \includegraphics[width=7cm,angle=0,clip]{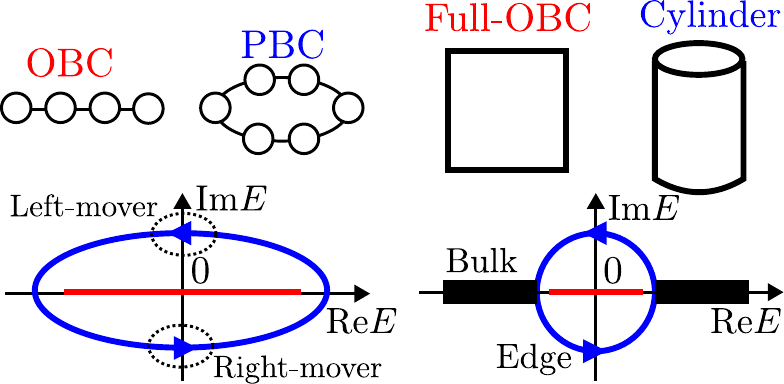}
 \caption{Schematic illustration of the non-Hermitian skin effects.
Left panel: Relationship between spectral winding and the skin effect.
Right panel: the skin effect on chiral edge states with edge-dependent dissipation.}
 \label{fig1}
\end{center}
\end{figure}

In recent years, the non-Hermitian skin effect—a phenomenon in which non-Hermitian lattice systems exhibit strong spectral sensitivity to boundary conditions and host boundary-localized skin modes—has attracted significant attention \cite{Hatano-Nelson-96,Hatano-Nelson-97,YW-18-SSH,zhang2022review, okuma2023non,lin2023topological,wang2024non,gohsrich2025non}.
In one-dimensional non-Hermitian tight-binding models, the appearance of the skin modes under the open boundary condition (OBC) is indicated by the nontrivial winding of the complex energy curve under periodic boundary conditions (PBC) (Fig. \ref{fig1}).
Related theorems about this correspondence were proven in Refs. \cite{OKSS-20,zhang-zhesen-chen}, following several observations made in Refs. \cite{Gong-18, Lee-Thomale-19, Borgnia-19}.
Moreover, based on the duality between Hermitian topological zero modes and non-Hermitian skin modes, Ref. \cite{OKSS-20} proposed the notion of symmetry-protected and higher-dimensional skin effects.

Anomaly-based interpretations \cite{lee2019topological,bessho2021nielsen} of non-Hermitian topology \cite{Gong-18,KSUS-19,zhou2019periodic} help to give an intuitive understanding of the non-Hermitian skin effect \cite{Okuma-Sato-21}.
Consider the PBC complex energy spectrum, where right-moving modes with a negative imaginary part and left-moving modes with a positive one coexist (Fig. \ref{fig1}).
In the corresponding OBC system, the right-moving and left-moving modes couple at the boundary, and the imaginary part of the resulting hybridized modes takes an intermediate value. 
A similar interpretation can be applied to more complex cases, including higher-dimensional skin effects \cite{okuma2023non,Okuma-Sato-21}.
These interpretations imply that various non-Hermitian skin effects can be induced by the combination of topological boundary states and boundary-dependent dissipation \cite{okuma2023non}.
For example, adding boundary-dependent dissipation to the chiral edge states of a Chern insulator, whose chirality depends on the boundary, leads to spectra similar to those observed in one-dimensional skin effects \cite{nakamura2023universal,schindler2023hermitian,ma2024non}.
In this case, the cylindrical and fully open boundary conditions (CBC and FOBC) correspond to the PBC and OBC for the edge states, respectively (Fig. \ref{fig1}).

Although the mathematical aspects have been extensively explored, uncovering the distinctive properties of the non-Hermitian skin effect in realistic quantum systems or solid-state settings remains one of the most challenging and intriguing problems in this rapidly growing field.
The bosonic Kitaev chain \cite{mcdonald2018phase,mcdonald2020exponentially} constitutes a novel platform for realizing the non-Hermitian skin effect. This effect does not stem from openness or dissipation, but from the intrinsic non-Hermiticity of the bosonic Bogoliubov–de Gennes (BdG) Hamiltonian, which fundamentally reflects the quantum nature of bosonic commutation relations.
Experimental studies have confirmed the validity of this mechanism \cite{slim2024optomechanical,busnaina2024quantum}.
A growing number of studies have theoretically investigated the non-Hermitian skin effect in more general bosonic BdG systems \cite{Flynn-20,Flynn-21,yokomizo2021non,okuma2022boundary,wang2022amplification,wan2023quantum,gomez2023driven,lv2024hidden}.

In this paper, we propose a platform to study the steady-state properties of the non-Hermitian skin effect, enabled by coupling bosonic topological edge states with BdG-type non-Hermiticity induced via parametric driving.
Unlike conventional models, our approach does not require complicated parametric driving terms.
Furthermore, the underlying bosonic topological systems have been extensively studied in various platforms, including tunable photonic systems \cite{Ozawa-review}, as well as magnonic \cite{mcclarty2022topological} and phononic \cite{liu2020topological} systems.

This paper is organized as follows.
In Sec. II, we briefly review the formalism of the bosonic BdG Hamiltonian and introduce the anomalous terms arising from parametric driving that will be used throughout this work.
In Sec. III, we present the model of bosonic chiral edge states under parametric driving and examine their spectral properties.
In Sec. IV, we compute the one-particle reduced density matrix defined from the nonequilibrium Green’s function, and show that a pronounced corner particle accumulation with quadrature anisotropy emerges in the steady state.
In Sec. V, we discuss the remaining topics and possible directions for future work.

\section{Bosonic BdG Hamiltonian under parametric driving} 
In this section, we briefly review the formalism of the bosonic BdG Hamiltonian. We also introduce the anomalous terms arising from parametric driving, which will be used throughout this paper.
\subsection{Bosonic BdG formalism}
Throughout this paper, we consider bosonic systems described by a BdG Hamiltonian \cite{bogoliubov1947theory,Altland-Simons,Colpa-78,kawaguchi2012spinor}: 
\begin{align}
    \hat{H}=\frac{1}{2}
    (\bm{a}^{\dagger},\bm{a}) H_{\rm BdG}
    \begin{pmatrix}
        \bm{a}\\
        \bm{a}^{\dagger}
    \end{pmatrix}
    =\frac{1}{2}(\bm{a}^{\dagger},\bm{a}) 
    \begin{pmatrix}
    h&s\\
    s^*&h^*
    \end{pmatrix}
    \begin{pmatrix}
        \bm{a}\\
        \bm{a}^{\dagger}
    \end{pmatrix},\label{bdgham}
\end{align}
where $(a_i,a_i^{\dagger})$ are the bosonic annihilation and creation operators on site $i$, $(\bm{a},\bm{a}^{\dagger})$ is the Nambu spinor, and $h$ and $s$ are Hermitian and symmetric matrices, respectively.
The terms $aa$ and $a^\dagger a^{\dagger}$ are called pairing terms or anomalous terms, which break particle-number conservation.
Unlike the case of fermions, the one-particle excitation energies cannot be obtained by diagonalizing the Hermitian matrix, $H_{\rm BdG}$.
The true energy spectrum is given by the eigenspectrum of the non-Hermitian matrix, $H_{\sigma\rm BdG}:=\sigma_z H_{\rm BdG}$, where $\sigma$'s denote the Pauli matrices acting on the Nambu space.
See Refs. \cite{Colpa-78,Gingras,Shindou-13,kawaguchi2012spinor} for details.
Here, the non-Hermiticity originates from the pairing (anomalous) terms, not from the open quantum nature.
The non-Hermitian Hamiltonian matrix $H_{\sigma\rm BdG}$ satisfies the following non-Hermitian symmetries \cite{BdGsym,KSUS-19}:
\begin{align}
    &\sigma_z[H_{\sigma\rm BdG}]^{\dagger}\sigma_z=H_{\sigma\rm BdG},\label{pseudo}\\
    &\sigma_x[H_{\sigma\rm BdG}]^{*}\sigma_x=-H_{\sigma\rm BdG}.\label{phs}
\end{align}
Equation (\ref{pseudo}) describes the pseudo-Hermiticity, which comes from the Hermiticity of $\hat{H}$.
Equation (\ref{phs}) is the bosonic analogy of the particle-hole symmetry.
Note that the corresponding Bloch Hamiltonian satisfies the following symmetries:
\begin{align}
&\sigma_z[H_{\sigma\rm{BdG}}(\bm{k})]^{\dagger}\sigma_z=H_{\sigma\rm{BdG}}(\bm{k}),\label{blochpseudo}\\
    &\sigma_x[H_{\sigma\rm{BdG}}(-\bm{k})]^{*}\sigma_x=-H_{\sigma\rm{BdG}}(\bm{k})\label{blochphs},
\end{align}
where $\bm{k}$ is the crystal momentum.

\subsection{Anomalous terms from parametric driving in rotating frame}
To introduce the anomalous terms, we consider the bosonic tight-binding model under the site-dependent parametric driving \cite{gerry2023introductory,peano2016topological}:
\begin{align}
    \hat{H}(t)=\sum_{i,j}a^{\dagger}_i\tilde{h}_{i,j}a_j+\sum_{i}\left[i\eta_ie^{i\Omega t}a_ia_i+h.c.\right],
\end{align}
where $\tilde{h}_{i,j}$ is the normal-part (unpaired) Hamiltonian matrix, $\eta_i$ characterizes the site-dependent anomalous term, and $\Omega$ is the driving frequency.
Typically, these anomalous terms arise from the nonlinear interactions such as $b^{\dagger}aa+h.c.$, where $b$ represents the bosonic pump modes that can be treated as classical coherent states. 
In the rotating frame, the time dependence of the Hamiltonian can be removed:
\begin{align}
    \hat{H}=\sum_{i,j}a^{\dagger}_ih_{i,j}a_j+\sum_{i}\left[i\eta_ia_ia_i+h.c.\right],\label{parametric}
\end{align}
where $h_{i,j}=\tilde{h}_{i,j}-(\Omega/2)\delta_{ij}$ is the renormalized normal part.
Mathematically, this is derived in the interaction picture by calculating $\hat{H}:=e^{i\hat{H}_0t}\hat{H}_1(t)e^{-i\hat{H}_0t}$ with $\hat{H}(t)=\hat{H_0}+\hat{H}_1(t)$ and $\hat{H}_0=\Omega/2\sum_{i}a^{\dagger}_ia_i$.
Clearly, Eq. (\ref{parametric}) can be rewritten in the form of Eq. (\ref{bdgham}).

\section{Spectral properties of skin effect in bosonic chiral edge states }
In this section, we introduce the model of bosonic chiral edge states under parametric driving that will be analyzed in this paper, and examine its spectral properties.
\subsection{Model}
In the following, we define the model analyzed in this paper.
As the renormalized normal part, we use the QWZ model \cite{Wu-Bernevig-Regnault-12}, whose Bloch Hamiltonian is given by
\begin{align}
    h(\bm{k})=\sin k_x\tau_x+\sin k_y\tau_y+(m-\cos k_x-\cos k_y)\tau_z,\label{qwz}
\end{align}
where $m$ denotes the mass term, and $\tau$'s are the Pauli matrices acting on the sublattice degrees of freedom within a unit cell.
For $0<m<2~(-2<m<0)$, the Chern number of the lowest band is 1 ($-1$).
Throughout this paper, we set $m=1$.
Thanks to the bulk-boundary correspondence \cite{hatsugai1993chern}, the chiral edge modes can be found under the OBC.
Note that the eigenvalues of the original Hamiltonian $\tilde{h}$ should be positive for thermodynamic stability, and the frequency of the parametric driving is chosen as the appropriate value to realize the Hamiltonian (\ref{qwz}).
The $\sigma$BdG form of the normal part is 
\begin{align}
    H^{(\rm{N})}_{\sigma\rm{BdG}}(\bm{k})&=\sin k_x\tau_x\otimes\hat{1}+\sin k_y\tau_y\otimes\sigma_z\notag\\
    &+(m-\cos k_x-\cos k_y)\tau_z\otimes\sigma_z,
\end{align}
where $\hat{1}$ is the identity matrix.
In the following, we omit the identity matrix  $\hat{1}$ whenever its inclusion is evident from the context.
The presence or absence of $\sigma_z$ in each term is determined by the particle-hole symmetry (\ref{blochphs}).
As the pairing term, we consider the following onsite Hamiltonian:
\begin{align}
    H^{(\rm{A})}_{\sigma\rm{BdG}}=\sum_{\bm{r}}i\Delta_{\bm{r}}\ket{\bm{r}}\bra{\bm{r}}\otimes\hat{1}\otimes\sigma_x,
\end{align}
where $\bm{r}$ denotes the unit cell, and $\Delta_{\bm{r}}\in\mathbb{R}$.
We assume that there are $L_x$ and $L_y$ unit cells along the $x$ and $y$ directions, respectively.
In addition to the symmetries (\ref{blochpseudo}) and (\ref{blochphs}), the total Hamiltonian $H_{\sigma\rm{BdG}}=H^{(\rm{N})}_{\sigma\rm{BdG}}+H^{(\rm{A})}_{\sigma\rm{BdG}}$ has the following onsite unitary symmetry:
\begin{align}
    [H_{\sigma\rm{BdG}},U]=0~~~~\mathrm{with}~U=\hat{1}\otimes \tau_x\otimes\sigma_x.\label{unitary-sym}
\end{align}
Thus, the Hamiltonian is block-diagonalized with respect to $U$, and each sector is characterized by the eigenvalues $U=\pm1$.
Below, we consider cases where this symmetry is important and where it is not. The former is significant as it provides an analogy to the bosonic Kitaev model \cite{mcdonald2018phase,mcdonald2020exponentially} in which the symmetry plays a crucial role, while the latter is important for considering a wide range of applications, including magnonic \cite{mcclarty2022topological} and phononic \cite{liu2020topological} systems.
In what follows, we begin by discussing the spectrum, and subsequently examine the properties of the nonequilibrium steady state.

\subsection{Spectral properties under edge driving}
\begin{figure}
\begin{center}
 \includegraphics[width=8.4cm,angle=0,clip]{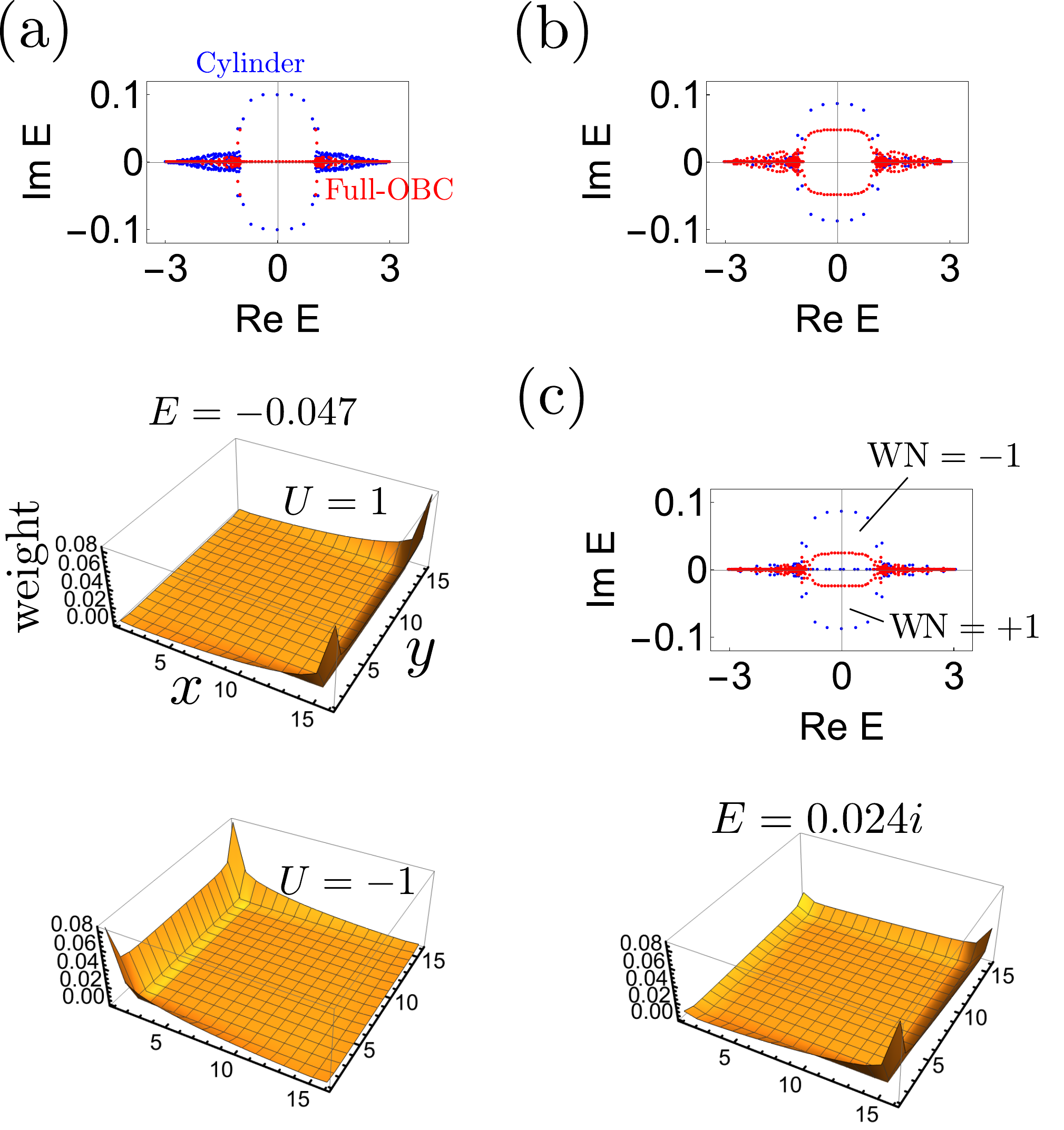}
 \caption{Non-Hermitian skin effect in bosonic chiral edge states under parametric driving. (a) Complex spectra and skin modes in the presence of the unitary symmetry (\ref{unitary-sym}). (b) Complex spectra under symmetry-breaking term. (c) Complex spectra and skin modes under one-edge driving at $y=1$.}
 \label{fig2}
\end{center}
\end{figure}

We consider the following two boundary conditions: the CBC and FOBC, both open in the $y$ direction. 
First, we assume uniform pairing terms on the bottom and top edges (i.e., $\Delta_{\bm{r}}=\Delta$
for $y=1$ and $y=L_y$, $\Delta_{\bm{r}}=0$ otherwise).
Under the CBC, by considering the symmetry \ref{unitary-sym}, we find that the low-energy effective theory at the edges is given by four chiral edge modes with complex energies:
\begin{align}
    (&k_x+ i \Delta)\ket{\tau_x=1}\bra{\tau_x=1}\otimes \ket{\sigma_x=1}\bra{\sigma_x=1},\label{pp}\\
    (&k_x- i \Delta)\ket{\tau_x=1}\bra{\tau_x=1}\otimes \ket{\sigma_x=-1}\bra{\sigma_x=-1},\label{pm}\\
    (-&k_x+ i \Delta)\ket{\tau_x=-1}\bra{\tau_x=-1}\otimes \ket{\sigma_x=1}\bra{\sigma_x=1},\label{mp}\\
    (-&k_x- i \Delta)\ket{\tau_x=-1}\bra{\tau_x=-1}\otimes \ket{\sigma_x=-1}\bra{\sigma_x=-1}.\label{mm}
\end{align}
Owing to the presence of complex energies with a positive imaginary part, this system exhibits dynamical instability, leading to a transition into another steady state.
In fact, the energy spectrum of the whole system with $L_x\times L_y=16\times16$ and $\Delta=0.1$ shows behavior explained by this effective theory [Fig. \ref{fig2}(a)].
A similar model with parametric driving at one edge was discussed in Ref. \cite{malz2019topological} in terms of topological magnon amplification.
In this sense, the essential contribution of this paper is in the FOBC case.
Under the FOBC, one can find the edge states with no imaginary part  [Fig. \ref{fig2}(a)], which means that the edge spectrum is completely different from the CBC one. This is a typical behavior of the non-Hermitian skin effect. Thanks to the skin effect, the system does not suffer from the dynamical instability of edge amplification.
Strictly speaking, a tiny imaginary part appears at the connection between edge states and bulk states. However, such small amplification may be masked by dissipation that should exist in a typical setup. This point will be discussed later in steady-state calculations.

In this skin effect, the unitary symmetry (\ref{unitary-sym}) plays a crucial role.
In the $U=1$ sector, there is one right mover with a positive imaginary part (\ref{pp}) and one left mover with a negative imaginary part (\ref{mm}).
These edge modes, together with bulk modes, form the spectral winding under the CBC, which signals the presence of the non-Hermitian skin effect.
Here, we have defined the one-dimensional winding number by y treating the $y$-direction as an internal degree of freedom.
In the $U=-1$ sector, the same phenomenon occurs with the opposite winding number.
Thus, the total winding number is zero.
In this sense, the above spectral behavior is understood as the skin effect protected by the unitary symmetry (\ref{unitary-sym}). In each sector, the skin modes are localized at $x=1$ or $x=L_x$ [Fig. \ref{fig2}(a)].
This means that the breakdown of the symmetry (\ref{unitary-sym}) can lead to the disappearance of the skin effect.
In our previous work \cite{Okuma-19}, this type of disappearance, known as infinitesimal instability, can be triggered by an infinitesimally small perturbation.
The same infinitesimal instability in the bosonic BdG system was also pointed out in Ref. \cite{yokomizo2021non}.
%However, in our model, which uses a one-dimensional subspace of a two-dimensional system, \textcolor{red}{the behavior of the spectrum becomes more complex.}
We add the symmetry-breaking term $\epsilon \hat{1}\otimes\hat{1}\otimes\sigma_z$ as a perturbation.
For $\epsilon=0.05$, we observe that the FOBC spectrum approaches the CBC spectrum. In the Appendix, we discuss this perturbative instability in more detail.
%Even for $\epsilon=0.05$, which is not so small with respect to $\Delta$, the FOBC spectrum is far from the CBC one [Fig. \ref{fig2}(b)].
%This phenomenon is intuitively understood by using the effective theory.
%In the absence of symmetry, the mode (\ref{pp}) can couple with the mode (\ref{mp}).
%However, these modes are localized at the opposite edges, and the local perturbation 
%does not connect these modes directly, unlike in the symmetry-protected skin effects in purely one-dimensional systems.
%Note that this spectral behavior is merely a remnant of the skin effect and is not protected by any non-Hermitian topological mechanism.

The symmetry specific to the model can be problematic for practical applications. 
If we do not require the energy of the skin modes to be real, the skin effect can be induced by applying parametric driving to only one edge $(y=1)$, even without the symmetry ($\epsilon=0.05$) [Fig. \ref{fig2}(c)].
This is because there exist two regions where the net spectral winding is $\pm1$.
By tuning the ratio between dissipation and $\Delta$, one can induce dynamical instabilities that depend on the boundary conditions.
Since this method does not rely on the specific details of the model, it is expected that the skin effect can be observed in a wide range of bosonic Chern insulators.

\begin{figure*}
\begin{center}
 \includegraphics[width=15.5cm,angle=0,clip]{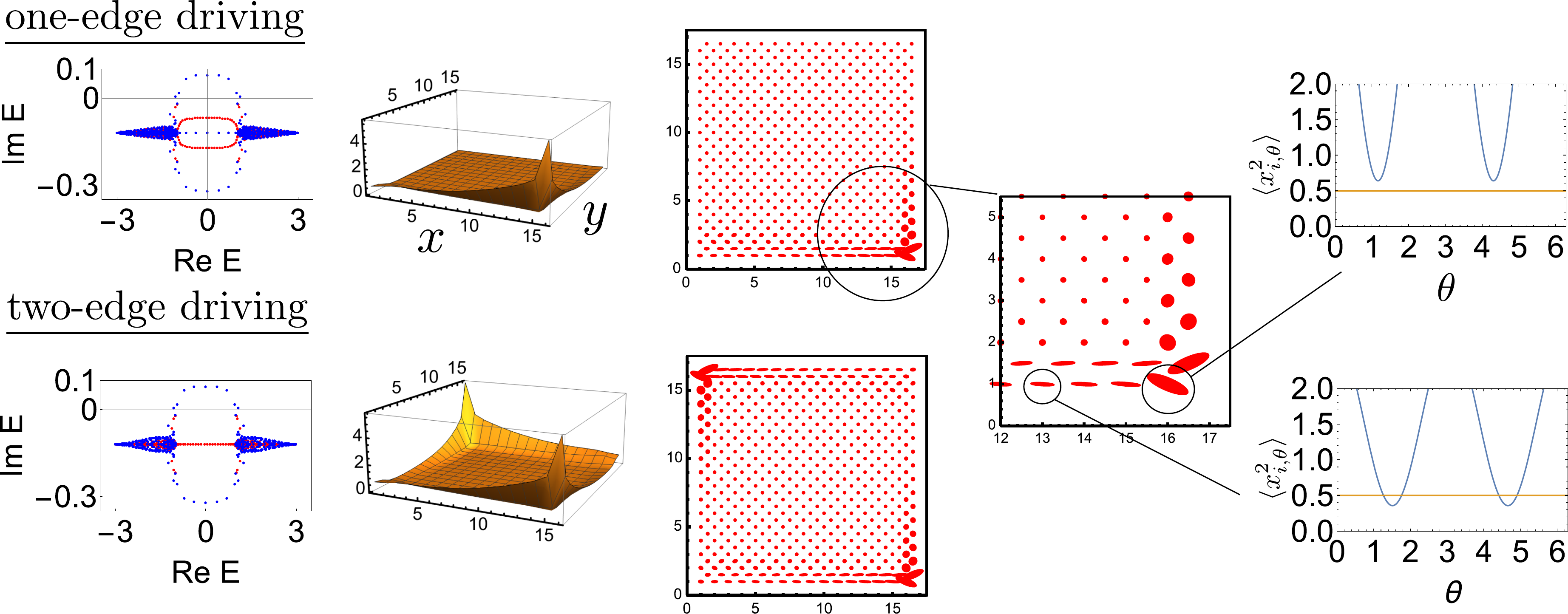}
 \caption{Steady-state accumulation caused by non-Hermitian skin effect. Left panel: Eigenspectra of the effective Hamiltonians and steady-state accumulation of $a$ bosons. Right panel: Spatial distribution of quadrature fluctuations, $\langle x^2_{i,\theta}\rangle$.
 The sublattice positions are plotted as (0,0) and (1/2,1/2) in each unit cell. For both one-edge and two-edge driving, the parameters are set as $L_x=L_y=16$, $\Delta=0.2$, $\epsilon=0$, and $\Gamma=0.12$.}
 \label{fig3}
\end{center}
\end{figure*}

\section{Steady-state accumulation with quadrature anisotropy}
In this section, we compute the one-particle reduced density matrix defined from the nonequilibrium Green’s function.
As a result, we find that a pronounced corner particle accumulation with quadrature anisotropy emerges.
\subsection{Self-energy from non-equilibrium nature}
In the absence of complex energies with a positive imaginary part, one can define the non-equilibrium steady state.
We define the $\omega$-dependent effective Hamiltonian by using the retarded Green's function in the Nambu representation \cite{okuma2022boundary}:
\begin{align}
    G^R(\omega)&=\frac{1}{G^{-1}_0-\Sigma^R_{\rm BdG}(\omega)}=\frac{1}{\omega\sigma_z-[H_{\rm BdG}+\Sigma_{\rm BdG}^R(\omega)]}\notag\\
    &=:\frac{1}{\omega-H^{(\rm eff)}_{\sigma\rm BdG}(\omega)}\sigma_z,\label{effective}
\end{align}
where $\Sigma^{R}_{\rm BdG}(\omega)$ is the retarded self-energy.
The $\omega\sigma_z$ term represents the difference from the fermionic case. 
In the presence of dissipation, while the effective Hamiltonian in $\sigma$BdG no longer has pseudo-Hermiticity, it still satisfies the following particle-hole symmetry \cite{okuma2022boundary}: 
\begin{align}
    \sigma_x [H^{(\rm eff)}_{\sigma\rm BdG}(-\omega)]^*\sigma_x=-H^{(\rm eff)}_{\sigma\rm BdG}(\omega).\label{omegadepphs}
\end{align}
To preserve the symmetry (\ref{unitary-sym}), we here consider the following $\omega$-independent dissipation:
\begin{align}
    \Sigma^R_{\sigma\rm{BdG}}(\omega):=\sigma_z\Sigma^R_{\rm{BdG}}(\omega)=-i\Gamma\hat{1}\otimes\hat{1}\otimes\hat{1},\label{dissipation}
\end{align}
where $\Gamma\in\mathbb{R}$. 
This type of dissipation preserves the mathematical structure of our model, except for a constant shift in the imaginary part of the complex energy. 
Our interest lies in FOBC systems whose corresponding CBC counterparts exhibit dynamical instability.

Note that the above dissipation (\ref{dissipation}) cannot be realized in thermal equilibrium, even under the approximation.
Suppose that dissipation is only in the normal part:
\begin{align}
    \Sigma^R_{\sigma\rm{BdG}}(\omega)=
    \begin{pmatrix}
        \sigma^R(\omega)&0\\
        0&-[\sigma^R(-\omega)]^*\\
    \end{pmatrix}.
\end{align}
For the thermal stability, $\sigma^R(\omega<0)=0$.
Thus, at a fixed $\omega\neq0$, the self-energy in thermal equilibrium becomes zero in either the particle or hole sector, which indicates that the dissipation (\ref{dissipation}) cannot be realized.
Under the parametric driving, the self-energy is shifted in the rotating frame:
\begin{align}
    \Sigma^R_{\sigma\rm{BdG}}(\omega)=
    \begin{pmatrix}
        \sigma^R(\omega+\frac{\Omega}{2})&0\\
        0&-[\sigma^R(-\omega+\frac{\Omega}{2})]^*\\
    \end{pmatrix}.\label{paradissipation}
\end{align}
If the experimentally relevant frequency is in $-\Omega/2<\omega<\Omega/2$, the self-energy (\ref{paradissipation}) does not contradict the dissipation (\ref{dissipation}) under approximation.
In summary, the approximation of imposing constant dissipation in a bosonic BdG system is physically meaningful only in nonequilibrium. This point is discussed, for example, in Refs. \cite{torre2013keldysh,sieberer2016keldysh}.

\subsection{steady-state accumulation}
To investigate the steady-state properties, we calculate the one-particle reduced density matrix.
The system bosons are assumed to be coupled to a thermal bath at a temperature sufficiently low compared to their excitation energies. That is, when the parametric pumping is turned off, the system bosons are not excited.
Under this assumption, the reduced density matrix in the rotating frame is given by \cite{torre2013keldysh,sieberer2016keldysh}:
\begin{align}
    \rho&=
    \begin{pmatrix}
        \langle\bm{a}^{\dagger}\bm{a}\rangle&\langle\bm{a}^{\dagger}\bm{a}^{\dagger}\rangle\\
        \langle\bm{a}\bm{a}\rangle&\langle\bm{a}\bm{a}^{\dagger}\rangle
    \end{pmatrix}
    \notag\\
    &=i\int \frac{d\omega}{2\pi}G^{<}(\omega)=\int \frac{d\omega}{2\pi}G^{\rm R}(\omega)
    \begin{pmatrix}
        0&0\\
        0&2\Gamma
    \end{pmatrix}
    G^{\rm A}(\omega),\label{densitymatrix}
\end{align}
where $G^<(\omega)$ is the lesser Green's function in non-equilibrium state, and the $2\times2$ matrix acts on Nambu space.
By using the formula (\ref{densitymatrix}), we calculate the steady-state quantities for both one-edge and two-edge driving.

First, we calculate the particle sector of $\rho$ and find an accumulation of $a$ bosons extremely localized at the corners (Fig. \ref{fig3}).
These results arise from transient amplification under strong non-normality of the effective Hamiltonian (\ref{effective}), characterized by the degree of non-commutativity of $[H,H^{\dagger}]$.
Under strong non-normality, the transient dynamics of linear systems are known to be governed by the $\epsilon$-pseudospectrum with a large imaginary part \cite{Trefethen}.
In our case, the wavepackets of bosons are amplified at one of the two edges and damped at the other edge.
Roughly speaking, in a system exhibiting the skin effect, the pseudospectrum for small $\epsilon$ is given by the two-dimensional region enclosed between the PBC and OBC curves in the complex plane \cite{okuma2023non}.
In the case of fermions, the PBC curve cannot have a positive imaginary part, and transient amplification does not occur.
As a result, the influence of the skin effect on transport remains limited compared to the bosonic BdG case.
The above considerations regarding the differences between bosons and fermions are also consistent with the results in Lindblad systems \cite{mcdonald2022nonequilibrium}.

Next, we consider the quadrature operator $x_{\theta}=(e^{i\theta}a+e^{-i\theta}a^{\dagger})/\sqrt{2}$, which is of particular interest in bosonic BdG systems \cite{torre2013keldysh,sieberer2016keldysh,gomez2023driven}. 
The $\theta=0$ and $3\pi/2$ cases correspond to the position and momentum operators $x$ and $p$ in harmonic oscillator systems, and to the spin operators $S_x$ and $S_y$ in magnonic systems. 
The quadrature fluctuation at each site $i$, $\langle x^2_{i,\theta}\rangle$, is calculated using Eq. (\ref{densitymatrix}).
In the vacuum state of $(a_i,a_i^{\dagger})$, $\langle x^2_{i,\theta}\rangle=1/2$, referred to as the standard quantum limit.
In Fig. \ref{fig3}, the magnitude and angle dependence of the quadrature fluctuations for steady-states are illustrated as an ellipse.
Since the two-dimensional bulk is almost in the vacuum state, the quadrature fluctuations are represented by a circle whose radius is given by the standard quantum limit.
On the parametrically driven edge, the system is affected by the skin effect, resulting in exponential localization of the particle number at the corner, along with an enhanced deformation of the quadrature ellipse.
Remarkably, the smallest value of the fluctuation is below the standard quantum limit at the edge, meaning that the steady state is squeezed [Fig. \ref{fig3}].
On the other hand, both the minor and major radii of the quadrature ellipse are greater than 1/2 at the corners [Fig. \ref{fig3}].
Nonetheless, the quadrature ellipse exhibits strong anisotropy, indicating that the particle accumulation in this steady state reflects the uniquely quantum nature characteristic of bosonic BdG systems.

\section{Discussion}
Before concluding the paper, we discuss several remaining issues.
An intriguing feature of this type of steady state is its sensitivity to small perturbations that connect the boundaries, which can lead to sharp changes in the steady-state density matrix $\rho$.
In the infinite-volume limit, even an infinitesimally small perturbation can cause the breakdown of the steady state. In our case, the split-ring geometry, consisting of a ring with a small gap, is well-suited for observing such instability.
Another key issue is the impact of boson-boson interactions.
Although their effects are negligible in photonic systems, it can play a significant role in quasi-bosonic systems such as magnonic platforms.
In a system of quasi-bosons defined emergently via a linear approximation, higher-order nonlinear terms act as interactions between the quasi-bosons.Whether the presence of such interactions leads to the suppression of exponential accumulation or to the emergence of a distance-dependent ordered state is a nontrivial question.
A concrete first step toward incorporating interactions would be to account for the frequency dependence of the self-energy in Eq. (\ref{effective}).
Overall, the physics of the bosonic skin effect represents a fascinating interplay among nonequilibrium dynamics, non-Hermiticity, interactions, and quantum effects.

\acknowledgements
I gratefully acknowledge valuable discussions with Chen-Hsuan Hsu and members of his group.
This work was supported by JSPS KAKENHI Grant No.~JP23K03243.
\\
\appendix
\section{Instability against symmetry-breaking term}
\begin{figure}
\begin{center}
 \includegraphics[width=8cm,angle=0,clip]{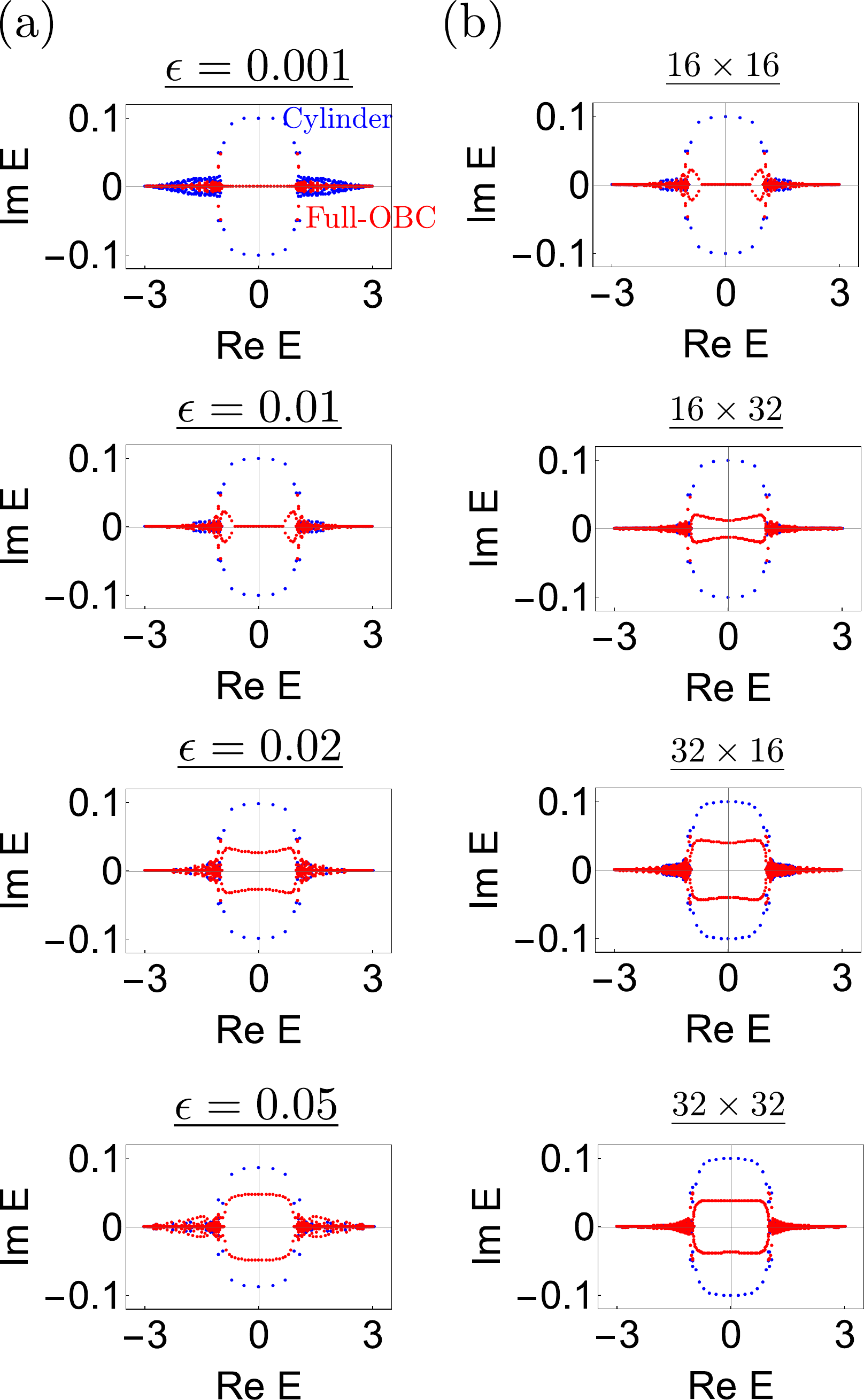}
 \caption{Complex energy spectra for different (a) perturbation strengths $(L_x\times L_y=16\times16)$ and (b) system sizes ($\epsilon=0.01$).}
 \label{supfig1}
\end{center}
\end{figure}

Figure \ref{supfig1} shows the spectra under small perturbations.
We find that for sufficiently large systems and sufficiently small perturbations, the spectrum does not change significantly. When the system size is fixed, the spectrum begins to change abruptly beyond a certain threshold value of the perturbation strength $\epsilon$.
Similarly, when $\epsilon$ is fixed, the spectrum becomes more sensitive to changes as the system size increases. These behaviors are qualitatively the same as those observed in the one-dimensional cases, such as symmetry-protected skin effects \cite{Okuma-19,OKSS-20} and critical skin effects \cite{li2020critical}.

On the other hand, increasing the system size while maintaining numerical accuracy is more difficult than in the one-dimensional case, and in the present numerical calculations, we were not able to study sufficiently large systems. Although the exact cause is unclear, even in the presence of perturbations, a discrepancy remains between the FOBC spectrum and the CBC spectrum.

\bibliography{NH-topo}

\end{document}